\documentclass[a4paper,11pt]{article}

\pdfoutput=1
 
\newcommand{\defaultlinespread}{\linespread{0.94}} 

\title{Connectivity of Natura 2000 forest sites in Europe}
\newcommand{\myTitle}{Connectivity of Natura 2000 forest sites in Europe}

\usepackage{geometry}
\defaultlinespread{}
\newenvironment{mtwocols}{\begin{multicols}{2}}{\end{multicols}}
 
\usepackage{array} 
\newcolumntype{L}[1]{>{\raggedright\let\newline\\\arraybackslash\hspace{0pt}}m{#1}}
\newcolumntype{C}[1]{>{\centering\let\newline\\\arraybackslash\hspace{0pt}}m{#1}}
\newcolumntype{R}[1]{>{\raggedleft\let\newline\\\arraybackslash\hspace{0pt}}m{#1}}

\newcolumntype{M}{>{\centering\arraybackslash$$}p{.985\linewidth}<{$$}}

\usepackage{adjustbox}%\usepackage{rotating}

\usepackage{environ}
\NewEnviron{mtvDisplayMath}{%
\vspace{4mm}
\noindent\colorbox{magenta!43!cyan!10!white}{%
\begin{tabular}{@{}M@{}} \\[-24mm]
\BODY \\[-1mm]
\end{tabular}}
\vspace{1mm}
}

\usepackage{amsmath}
\usepackage{amsfonts} % \Box

\usepackage{authblk}

\usepackage[T1]{fontenc}
\usepackage{fourier}

\usepackage{fancyhdr}
\usepackage{forloop}
\usepackage[usenames,dvipsnames]{color}
\usepackage[hyphens]{url}
\usepackage{hyphenat} 
\usepackage[font=footnotesize,labelfont=bf]{caption} 

\usepackage[pagebackref=true]{hyperref}
\definecolor{Brown}{rgb}{0.61,0.08,0.23} 
\definecolor{Blues}{rgb}{.12,.12,.48}
\hypersetup{
  colorlinks   = true,  %Colours links instead of ugly boxes
  urlcolor     = Blues, %Colour for external hyperlinks
  linkcolor    = Brown, %Colour of internal links
  citecolor    = Brown  %Colour of citations
}

\usepackage{cite}
\usepackage{multicol}
\usepackage{fancybox}
\usepackage{tikz}
\usepackage{float}

\usepackage{enumitem}

\usepackage{array}
\newcolumntype{L}[1]{>{\raggedright\let\newline\\\arraybackslash\hspace{0pt}}m{#1}}
\newcolumntype{C}[1]{>{\centering\let\newline\\\arraybackslash\hspace{0pt}}m{#1}}
\newcolumntype{R}[1]{>{\raggedleft\let\newline\\\arraybackslash\hspace{0pt}}m{#1}}

\newcommand{\checkis}[1]{\href{http://mastrave.org/doc/mtv\_m/check\_is\#SAP\_#1}{\bf{::#1::}}}

\newcommand{\gofrom}[2]{}

\newcommand{\gourl}[1]{\url{#1}}
\newcommand{\goDOI}[1]{\href{http://dx.doi.org/#1}{DOI:#1}}

\newcommand{\partitle}[1]{\vspace{3mm}\noindent\colorbox{black!10}{\begin{tabular}{C{64mm}}{\textbf{\textsf{\sloppy #1}}}%\\[-0.5mm]
\end{tabular}}\\[0mm]}
\usepackage{titlesec}
\titleformat*{\section}{\Large\bf\sffamily}

\newcounter{mylabelnumber}

\newcommand{\checkendpaper}{
\setlength{\leftmargin}{0pt}
\setlength{\itemindent}{0em}

\setlength{\itemsep}{0mm}
\setlength{\parskip}{0mm}
\setlength{\parsep}{0mm} 
\linespread{0.92} 
}

\renewcommand*{\backref}[1]{}
    \renewcommand*{\backrefalt}[4]{%
    \ifcase #1 %
        (Not cited).
    \or
        (page~#2).%
    \else
        (pages~#2).
    \fi}

\usepackage{lettrine} 
\usepackage{colortbl} % Initial info-table

\makeatletter
\def\newparshape{\parshape\@npshape0{}}
\def\@npshape#1#2#3{\ifx\\#3\expandafter\@@@npshape\else\expandafter\@@npshape\fi
{#1}{#2}{#3}}
\def\@@npshape#1#2#3#4#5{%
\ifnum#3>\z@\expandafter\@firstoftwo\else\expandafter\@secondoftwo\fi
{\expandafter\@@npshape\expandafter{\the\numexpr#1+1\relax}{#2 #4 #5}{\numexpr#3-1\relax}{#4}{#5}}%
{\@npshape{#1}{#2}}}
\def\@@@npshape#1#2#3{#1 #2 }
\makeatother

\title{\vspace{-24mm}\bf \huge \textsf{\myTitle{}}\vspace{3mm}}
\author[1,\textbf{*}]{{\Large \textsf{Christine Estreguil}} \vspace{2mm}}
\author[1]{{\Large \textsf{Giovanni Caudullo}} \vspace{2mm}}
\author[1,2]{{\Large \textsf{Daniele de Rigo}} \vspace{-3mm}}

\affil[1]{\small \;European Commission, Joint Research Centre, Institute for Environment and Sustainability \\

Via E. Fermi 2749, I-21027 Ispra (VA), Italy\smallskip }
\affil[2]{\;Politecnico di Milano, Dipartimento di Elettronica, Informazione e Bioingegneria\\

Via Ponzio 34/5, I-20133 Milano, Italy\vspace{-4mm}}
\date{}
\begin{document}

  \maketitle
  
\noindent\colorbox{black!10}{\small
\arrayrulecolor{white}\color{black!80}\begin{tabular}{|p{138mm}|}
\hline
\vspace{0mm}
\textbf{Background/Purpose}: In the context of the European Biodiversity policy, the Green Infrastructure Strategy is one supporting tool to mitigate fragmentation, inter-alia to increase the spatial and functional connectivity between protected and unprotected areas. The Joint Research Centre has developed an integrated model to provide a macro-scale set of indices to evaluate the connectivity of the Natura 2000 network, which forms the backbone of a Green Infrastructure for Europe. The model allows a wide assessment and comparison to be performed across countries in terms of structural (spatially connected or isolated sites) and functional connectivity (least-cost distances between sites influenced by distribution, distance and land cover).\\[2mm]

\textbf{Main conclusion:} The Natura 2000 network in Europe shows differences among countries in terms of the sizes and numbers of sites, their distribution as well as distances between sites. Connectivity has been assessed on the basis of a 500 m average inter-site distance, roads and intensive land use as barrier effects as well as the presence of "green" corridors. In all countries the Natura 2000 network is mostly made of sites which are not physically connected. Highest functional connectivity values are found for Spain, Slovakia, Romania and Bulgaria. The more natural landscape in Sweden and Finland does not result in high inter-site network connectivity due to large inter-site distances. The distribution of subnets with respect to roads explains the higher share of isolated subnets in Portugal than in Belgium.
\vspace{5mm}

Cite as:\\[2pt]
Estreguil, C., Caudullo, G., de Rigo, D., 2014. {\bf \myTitle{}}. \href{http://f1000.com/posters/browse/summary/1095638}{{\em F1000Posters 2014, 5}: 485}. DOI: \href{http://dx.doi.org/10.6084/m9.figshare.1063300}{10.6084/m9.figshare.1063300}. ArXiv: \href{http://arxiv.org/abs/1406.1501}{1406.1501}
\vspace{5mm}

\textbf{* Corresponding author}: christine.estreguil@jrc.ec.europa.eu
\\[1mm]
\hline
\end{tabular}}

%-------------------------------------
%%%%%%%% SECTION 1
%-------------------------------------

\vspace{5mm}\begin{mtwocols} 
\noindent \lettrine[lines=2]{T}{he} Europe's Green Infrastructure Strategy is a holistic policy initiative integrating nature, biodiversity and sustainable development \cite{Bennett_etal_2010}; it is one supporting tool to implement the European Biodiversity Strategy and achieve 2020 targets \cite{EC_2011}. One of its aims is to mitigate fragmentation and render protection more effective. One condition to achieve this is to increase the spatial and functional connectivity between natural and semi-natural protected and unprotected areas.

Protected areas such as Natura 2000 (N2K) \cite{EC_1992} sites form the backbone of a Green Infrastructure for Europe \cite{DGENV_2012a, EC_2013}. Besides their size and quality, their connectivity contributes to the movement and dispersal of animals and plants. 

\end{mtwocols}
\begin{figure}[H]

\vspace{-1mm} 
\centerline{\includegraphics[width=13cm]{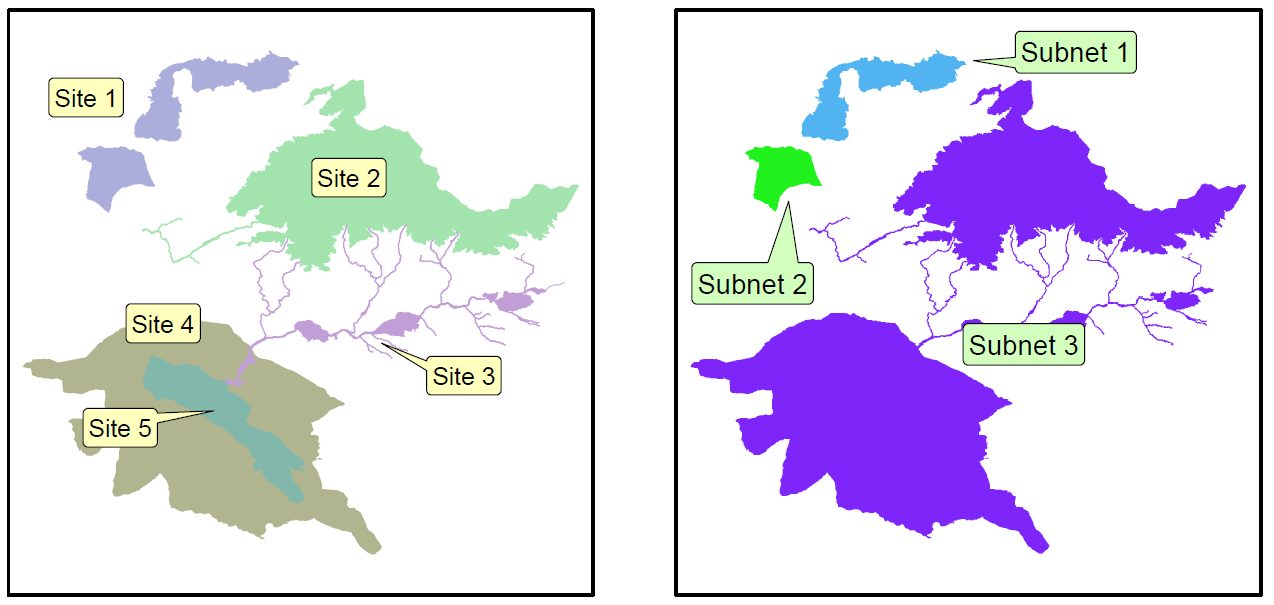}}
\vspace{-2mm} 
\caption{Illustration of the site rasterization method: five N2K sites (left) resulting three subnets (right, from~\cite{Estreguil_etal_2013}).}
\end{figure}
\begin{mtwocols} 

The unprotected landscape plays a role in enhancing or reducing the conservation and resilience of protected habitats; for example, grey infrastructure (artificial lands and roads) and intensive land use often pose the biggest threats or disturbances for biodiversity conservation. There is a need for conservation tools accounting for connectivity in tandem with landscape planning~\cite{McHugh_Thompson_2011}.

Connectivity measures should not necessarily be to link individual habitat patches with physical structures (such as corridors of similar habitat), but to ensure the existence of required \textbf{functional connections between protected sites} (e.g. inter-site \textbf{distances} or/and \textbf{landscape permeability}) and identify potentially isolated sites.

The Joint Research Centre has developed an integrated spatially explicit model to provide a macro-scale European vision of the connectivity of N2K sites on\cite{Estreguil_etal_2013, Estreguil_etal_2014}:
\begin{itemize}[leftmargin=*, noitemsep, topsep=1pt]
\item \textit{Site network pattern} in terms of size and number.
\item \textit{Intra-site network structural connectivity} with morphological criteria: complex, simple subnets.
\item \textit{Inter-site network functional connectivity} from the perspective of influencing factors related to site distribution, grey infrastructure and inter-site distance: connectivity indices, isolated sites.
\end{itemize}

%-------------------------------------
%%%%%%%% SECTION 2
%-------------------------------------
\partitle{Pattern of Natura 2000 site network}

\noindent The network of Natura 2000 sites designated under the "Habitat"\cite{EC_1992} and "Birds" Directives \cite{EC_2010} covers circa 18\% of the European Union (728,744 $\text{km}^{\text{2}}$); 80\% of the sites include forest. All N2K polygons\cite{DGENV_2012b} representing the extracted site areas have been converted to a raster layer in order to generate sub-network (subnet) components which were formed by one or more Natura 2000 sites in cases of overlap (sites physically connected) (Figure 1).

There are differences among countries in terms of the sizes and numbers of sites, their distribution as well as distances between sites (Figure 2). In the small frames of Figure 2, Ireland (frame 1) has a well-connected network; Portugal (frame 2) has large and isolated subnets; Spain (frame 3) has large and connected sites; Germany (frame 4) has small and densely distributed sites; Sweden (frame 5) has mainly small and distant subnets; Bulgaria (frame 6) has large subnets.

\end{mtwocols}

\begin{figure}[H]
\vspace{25mm}\begin{adjustbox}{addcode={\begin{minipage}{\width}}{\caption{%
      Natura 2000 forest site network with six zoom frames showing network differences.
      }\end{minipage}},rotate=90,center}
      \centerline{\includegraphics[height=13.8cm]{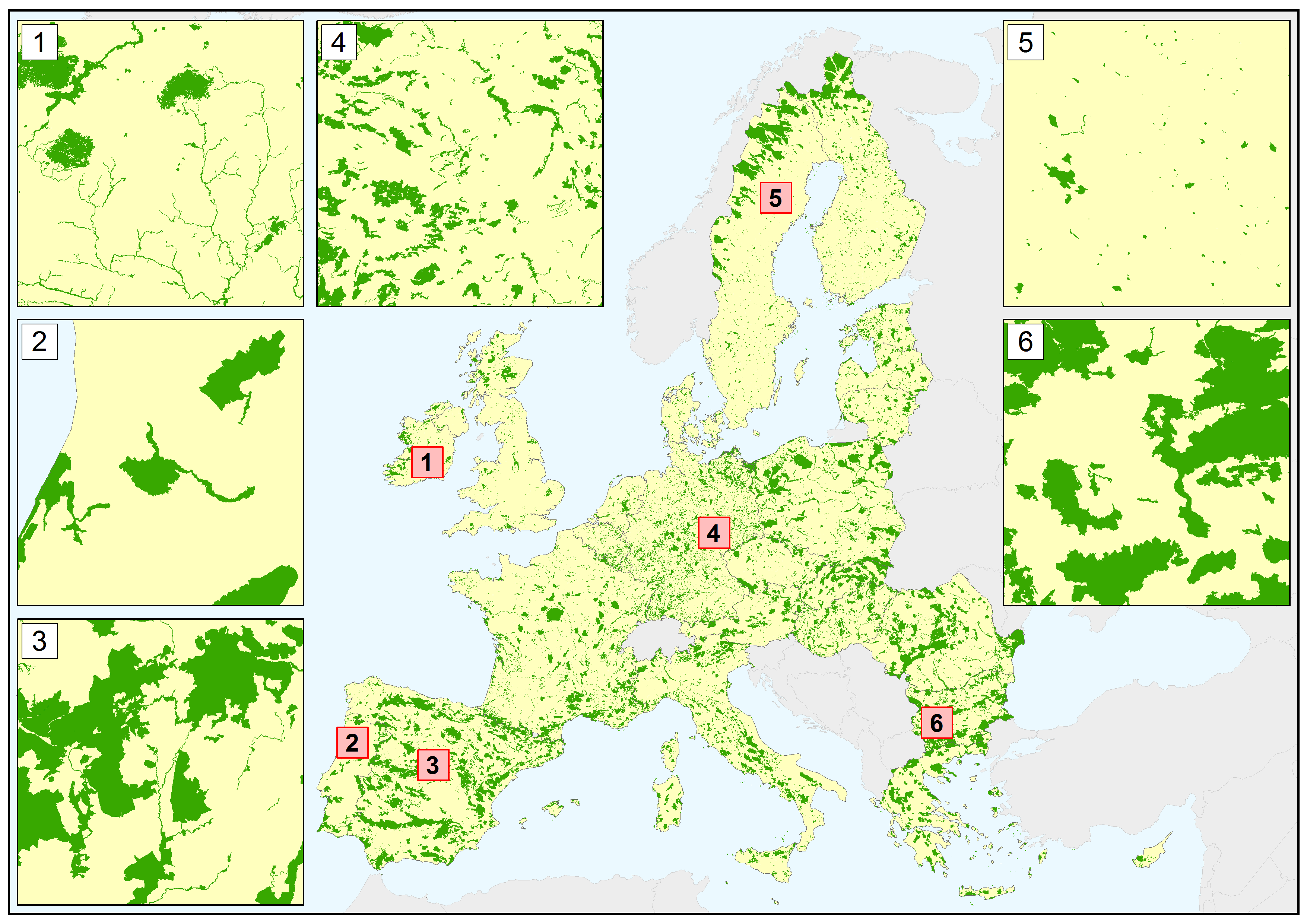}}
\end{adjustbox}
\end{figure}
\begin{mtwocols} 
\bigskip

\end{mtwocols}
\begin{figure}[H]

\vspace{20mm} 
\begin{adjustbox}{addcode={\begin{minipage}{1.05\width}}{\caption{%
      The data input and information flow to compute the analysis on Natura 2000 network connectivity. The intermediate steps generate \checkis{categorical} raster \checkis{matrix} layers from the input data. 
      The categorical variables are then transformed into \checkis{nonnegative} friction maps and functional distance (cost) estimates, from which least-cost paths are computed. Most derived indices are dimensionless \checkis{proportion} values in $[0,1]$ with the exception of the \checkis{nonnegative} indices summarising the number of N2K sites and of subnets, and their size. The notation in the workflow follows the Semantic Array Programming paradigm as applied in \cite{Estreguil_etal_2012,Estreguil_etal_2014} where the array-based semantics is based on \cite{de_Rigo_2012}. Details on the less obvious data-transformation modules may be found in \cite{Estreguil_etal_2014b}.
      }\end{minipage}},rotate=90,center}
      \centerline{\includegraphics[height=12cm]{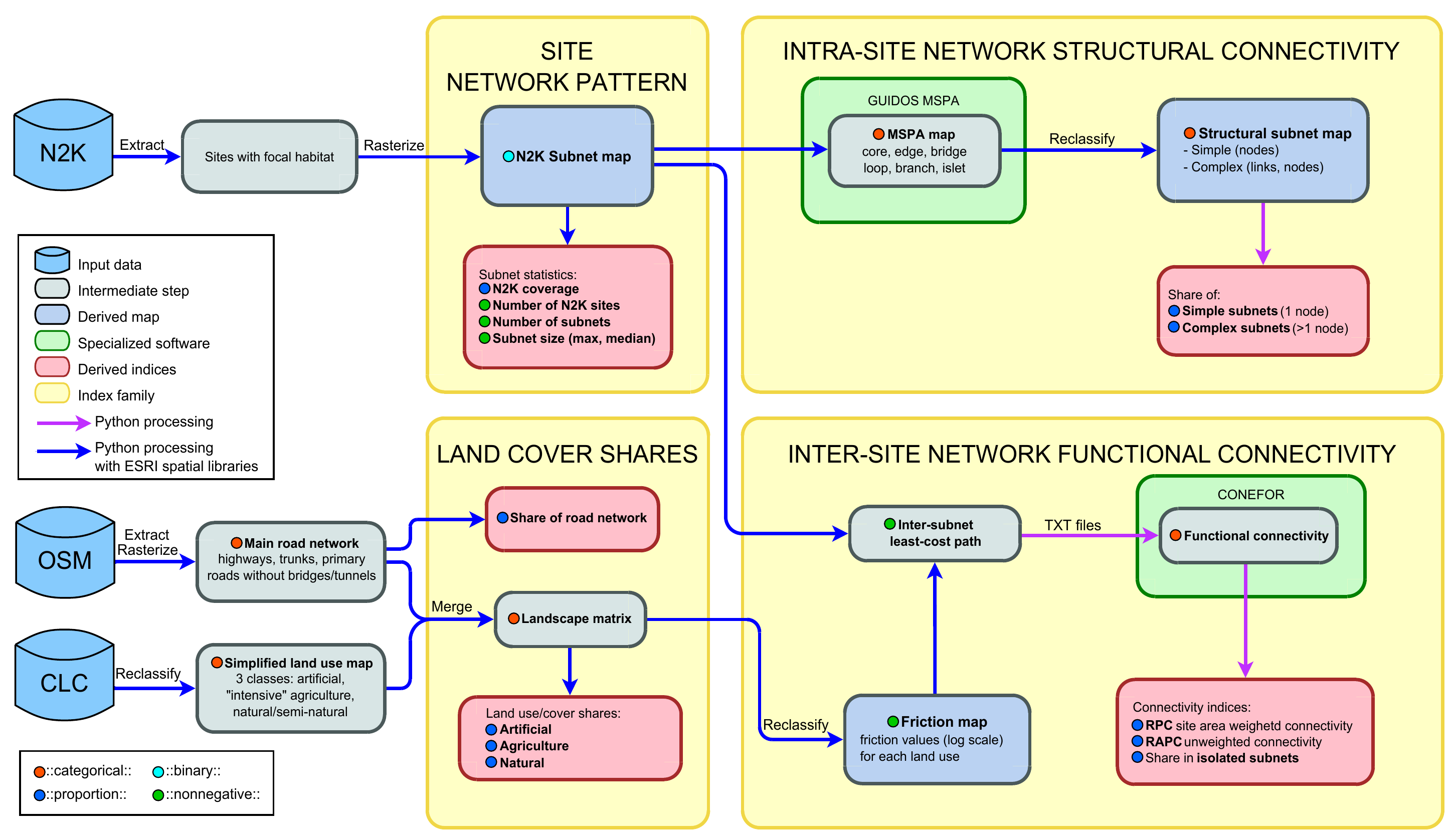}}%
\end{adjustbox}

\end{figure}
\begin{mtwocols} 
\bigskip

\end{mtwocols}
\begin{figure}[H]
\vspace{-1mm} 
\centerline{\includegraphics[width=13cm]{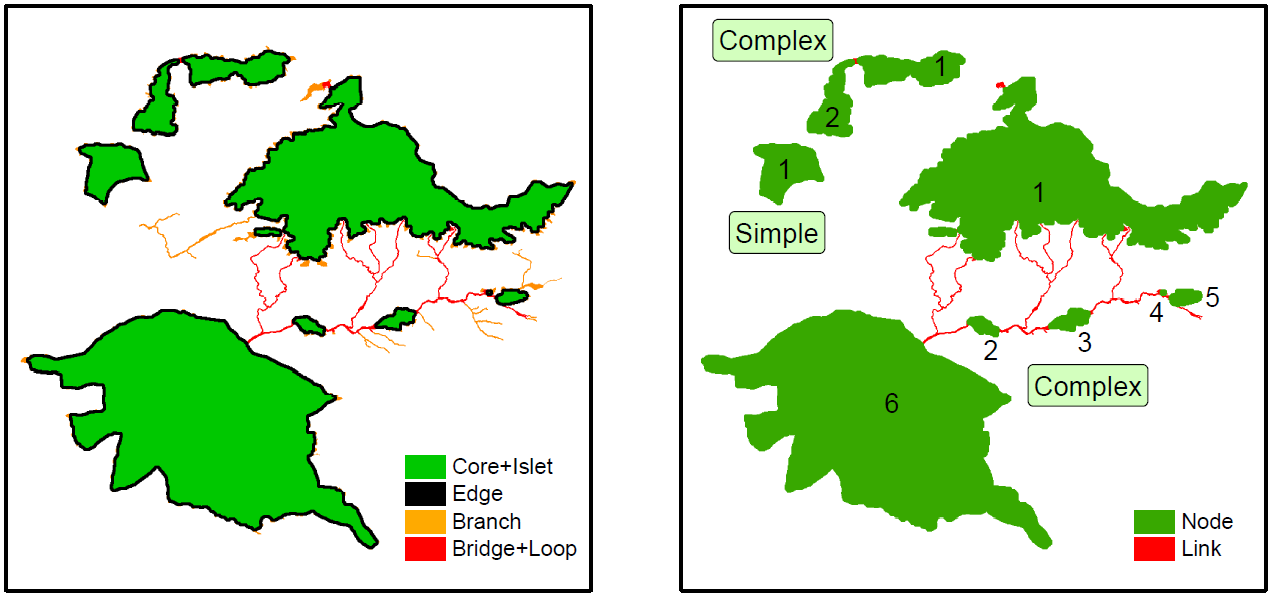}}
\vspace{-2mm}
\caption{Structural connectivity morphological analysis of subnets based on GUIDOS software (left), re-classification into simple/complex subnets (right, from \cite{Estreguil_etal_2013}).}
\end{figure}
\begin{mtwocols} 
\smallskip

%-------------------------------------
%%%%%%%% SECTION 3
%-------------------------------------
\partitle{Spatially-explicit model of connectivity of protected areas}

\noindent The model is based on two available software packages (free-download GUIDOS\cite{Soille_Vogt_2009} and free software Conefor\cite{Saura_Torne_2009}) integrated with GIS Python programming tools for automated processing\cite{Van_Rossum_Drake_2011} (Figure 3). The approach is harmonized and applies both \textbf{structural} and \textbf{functional} criteria. It represents a compromise between biological species models and commonly used connectedness measures.

%-------------------------------------
%%%%%%%% SECTION 4
%-------------------------------------
\partitle{Intra-site network structural connectivity}

\noindent The spatial configuration of the sites is characterized in terms of simple subnets, made of one node, and of complex subnets, made of several interconnected nodes and links  (Figure 4).
 For each country two structural connectivity indices are proposed:
\begin{itemize}[leftmargin=*, noitemsep, topsep=1pt]
\item Share of \textit{Complex subnets}.
\item Share of \textit{Simple subnets}.
\end{itemize}

%-------------------------------------
%%%%%%%% SECTION 5
%-------------------------------------
\partitle{Inter-site network functional connectivity}

\noindent The model is based on a probabilistic power weighted dispersal function \cite{Estreguil_etal_2014}. As a proxy of landscape resistance to species dispersal and to identify "green" corridors, a new European land use based friction map was created from the Corine Land Cover map of year 2006 at 100 m spatial resolution \cite{EEA_2012} and the OpenStreetMap layer\cite{Haklay_Weber_2008, Bennett_2010}.

\begin{figure}[H]
\vspace{-1mm} 
\centerline{\includegraphics[scale=0.30]{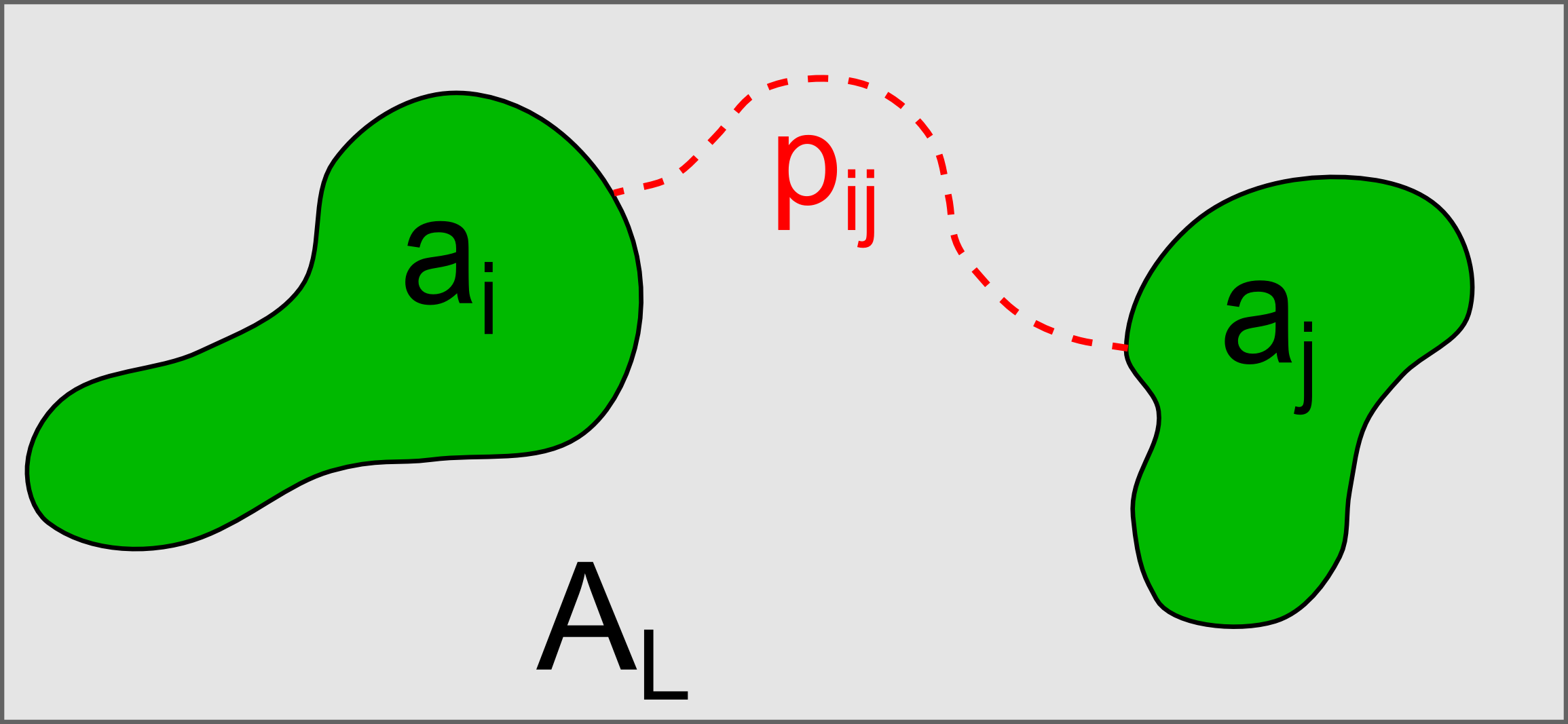}}
\vspace{-2mm}
\caption{Illustration of functional connectivity and parameters: $a_{i}$ and $a_{j}$ refer to subnet areas, $p_{i,j}$ is the probability of connectivity, $A_{L}$ the area of the landscape unit.}
\end{figure}

\noindent The probability of connectivity ($\text{\em p}_{ij}$) is measured as a function of the landscape resistance ($\text{\em cost}_{ij}$) between subnets ($\text{\em a}_{i}$, $\text{\em a}_{j}$). It applies 50\% probability of connectivity at 500 m average inter-subnets distance ($\text{\em dist}_{\,50\%}$). It also accounts for the presence of "green" corridors in-between subnets (Figure 5).

\[
\begin{array}{lll}
   \text{\em k} &=& \displaystyle \frac{\operatorname{ln}(0.5)}{dist_{\,50\%}} \\[5mm]
   p_{i,j}                &=& \displaystyle e^{\,\text{\em k} \,\cdot\, \text{\em cost}_{i,j}}
\end{array}
\]

\end{mtwocols}
\begin{figure}[H]
\vspace{-1mm} 
\centerline{\includegraphics[width=15cm]{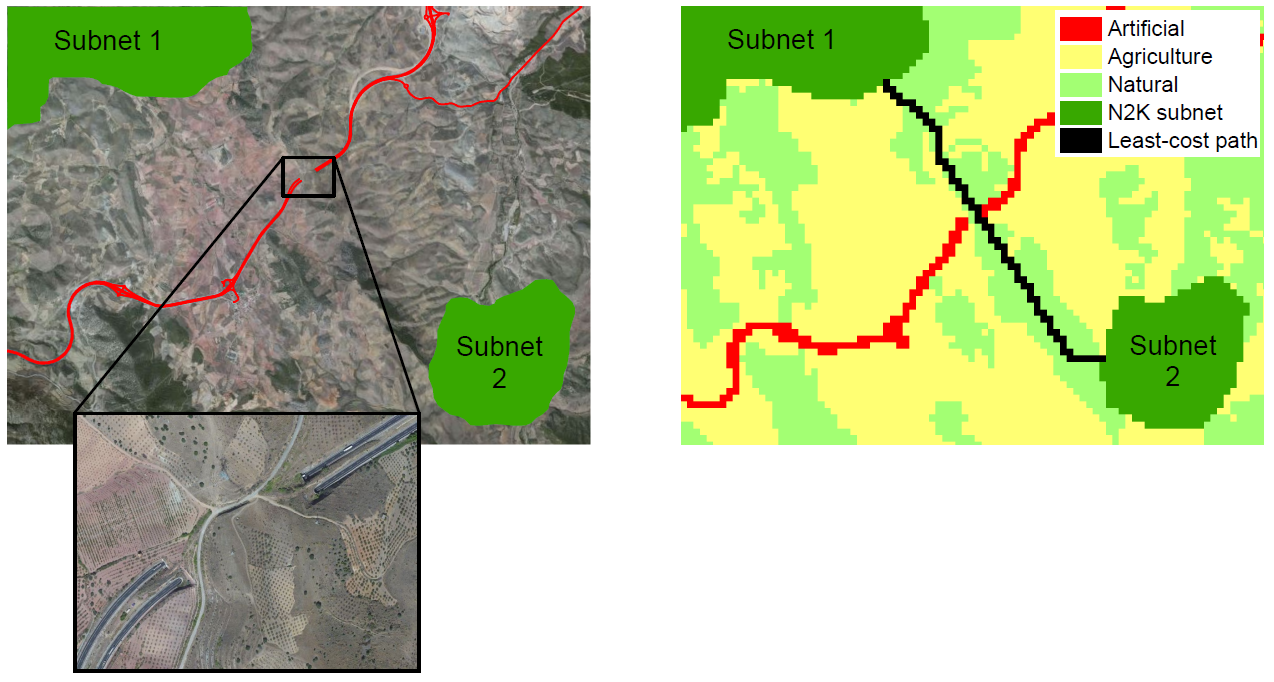}}
\vspace{-2mm}
\caption{Functional connectivity illustrated for two subnets; the tunnel along the main road (left) is accounted for identifying the least-cost (from \cite{Estreguil_etal_2013}).}
\end{figure}
\begin{mtwocols} 

\end{mtwocols}
\begin{table}[h]
\vspace{-2mm}
\centerline{\includegraphics[width=15cm]{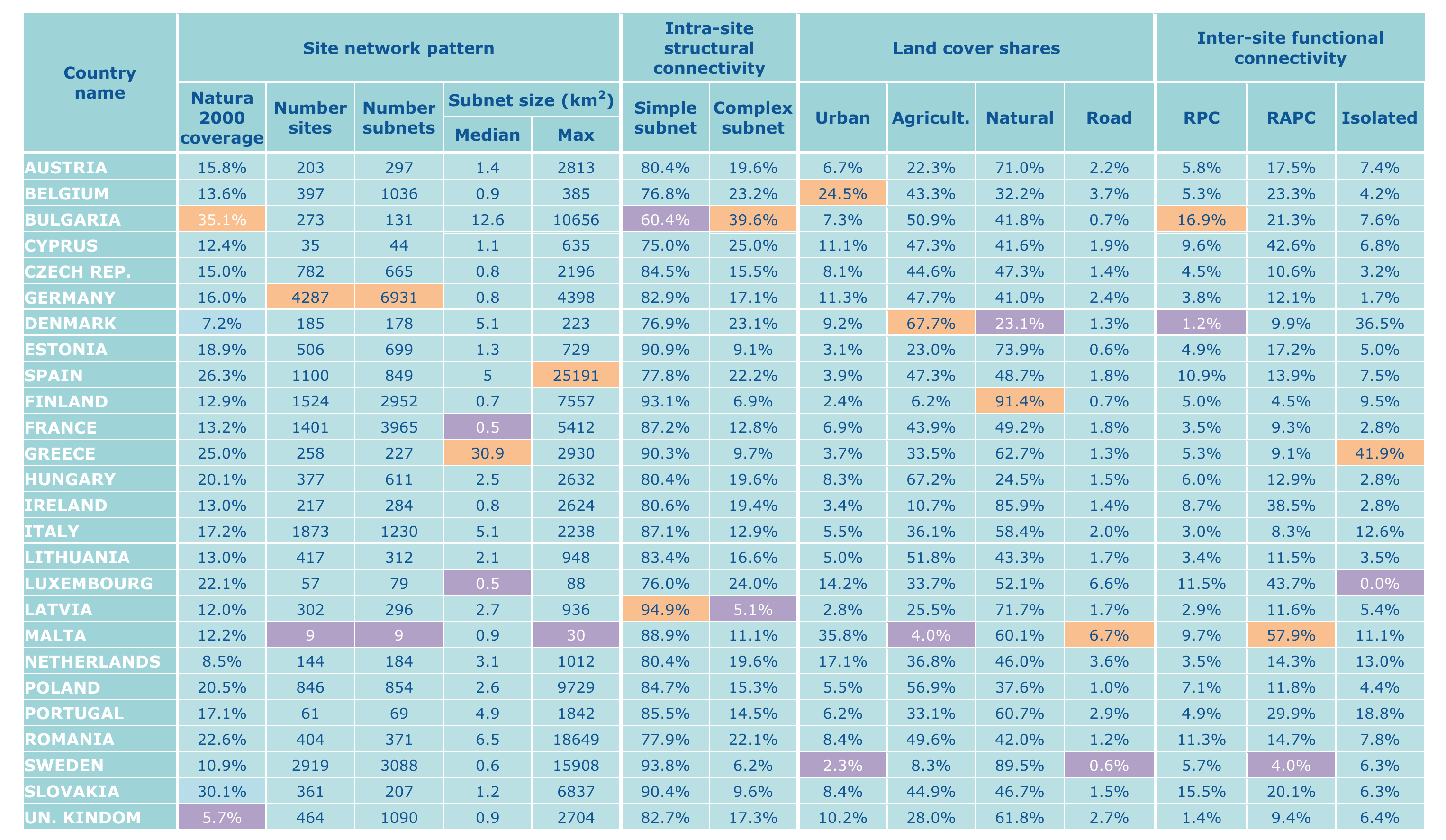}}
\vspace{-2mm}
\caption{Country based table of indices highlighting their highest value (orange) and their lowest (purple).}
\end{table}
\begin{mtwocols}

\noindent For each country three functional connectivity indices are applied to the subnets\footnote{\textit{RPC} and \textit{RAPC} are special cases of the general family of indices \textit{Power Weighted Probability of Dispersal} (PWPD) \cite{Estreguil_etal_2014}. They also belong to its simplified formulation s-PWPD \cite{Estreguil_etal_2012} as instances in which the s-PWPD parameters $\{\alpha, \beta, \gamma_1, \gamma_2\}$ are respectively $\{0, 1/2, 1, 1\}$ for RPC and $\{0, 1/2, 0, 0\}$ for RAPC.}:
\begin{itemize}[leftmargin=*, noitemsep, topsep=1pt]
\item \textit{Site area weighted Root Probability of Connectivity} (RPC) which is sensitive to the size of subnets. 
\[
   \text{\em RPC} \quad = \quad \sqrt{\frac{\displaystyle \sum_{i=1}^n \sum_{j=1}^n a_i \cdot a_j \cdot p_{ij}}{A_{L}^2}}
\]

\item \textit{Root un-weighted Average Probability of Connectivity} (RAPC) which is sensitive to the unprotected landscape resistance and functional distances between subnets.
\[
   \text{\em RAPC} \quad = \quad \sqrt{\frac{\displaystyle \sum_{i=1}^n \sum_{j=1}^n p_{ij}}{n^2}}
\]

\item Share of functionally \textit{Isolated subnets} in the Natura 2000 network.
\end{itemize}

\bigskip

%-------------------------------------
%%%%%%%% SECTION 6
%-------------------------------------
\partitle{Country based results on the Natura 2000 network connectivity}

\noindent In all countries, the N2K network is mostly made of simple subnets. Share of complex subnets ("physically connected" sites) range from 40\% in Bulgaria to 5\% in Latvia.  

Large numbers of subnets and similar small median sizes are found in Germany and France, their inter-site connectivity with emphasis on the sites' sizes is rather low (RPC approx. 3\%) but higher when the inter-site distances and landscape are at focus (RAPC approx. 10\%).

Highest RPC are found for Spain, Slovakia, Romania and Bulgaria. The more natural landscape in Sweden and Finland does not result in high inter-site network connectivity due to large inter-site distances. The distribution of subnets with respect to roads explains the higher share of isolated subnets in Portugal than in Belgium.

\end{mtwocols}
\begin{figure}[H]
\vspace{-1mm} 
\centerline{\includegraphics[width=7.1cm]{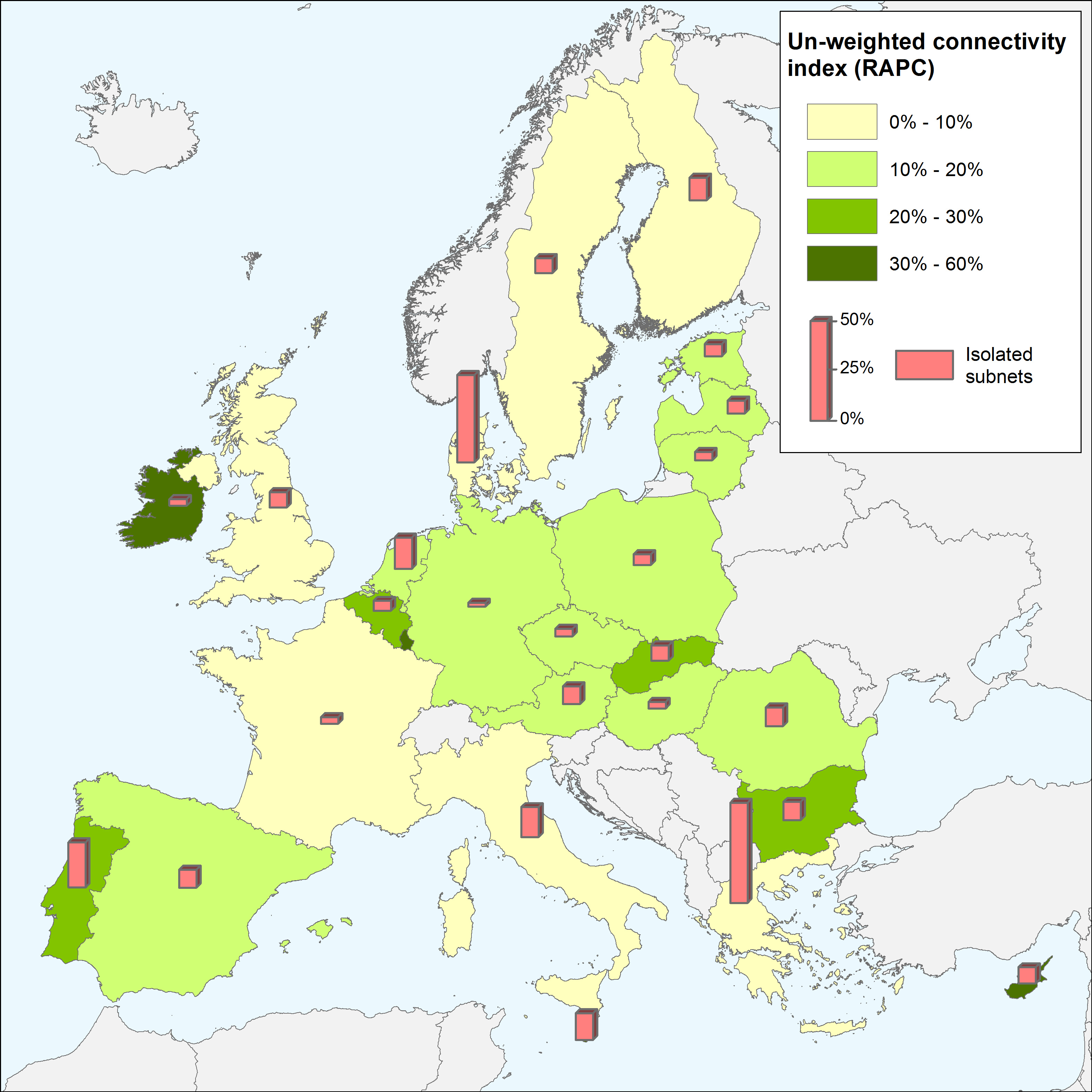}}
\vspace{-2mm}
\caption{European map of the root un-weighted average probability of connectivity (RAPC) and the share of N2K subnets functionally isolated.}
\vspace{-1mm} 
\end{figure}

\begin{figure}[H]
\vspace{0mm}
\centerline{\includegraphics[width=14cm]{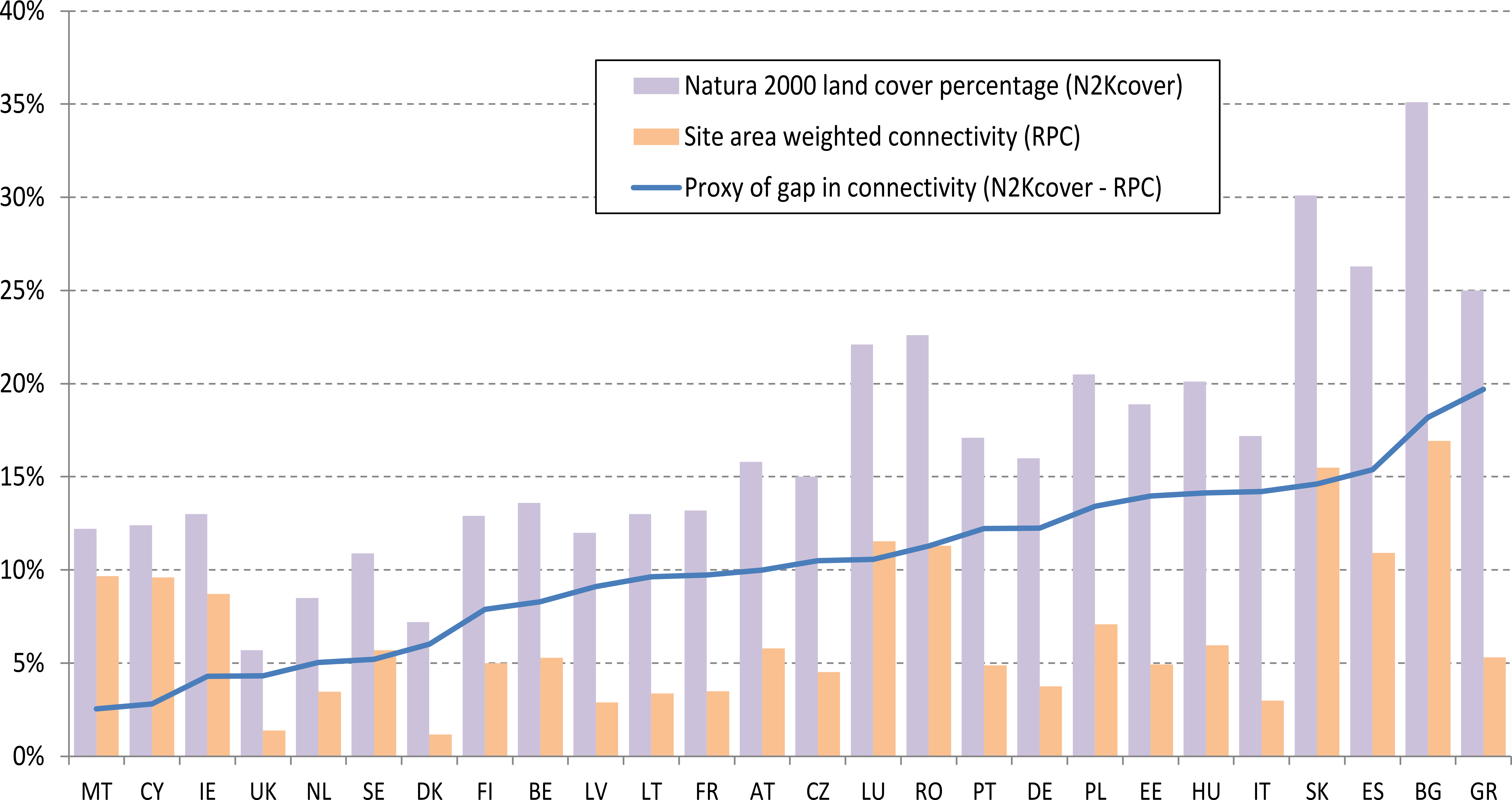}}
\vspace{-2mm}
\caption{National profile of site area weighted root probability of connectivity (RPC). The chart includes the N2K cover percentage per country and a proxy of the gap in connectivity computed as their difference (from~\cite{Estreguil_etal_2013}).}
\end{figure}

\vspace{8mm}
\noindent\textbf{For further datails please refer to:}
\smallskip

\noindent Estreguil, C., Caudullo, G., San-Miguel-Ayanz, J., 2013. Connectivity of Natura 2000 Forest Sites. EUR 26087EN. Luxemburg: Publications Office of the European Union. JRC 83104. \goDOI{10.2788/95065}
\smallskip

\noindent Estreguil, C., de Rigo, D., Caudullo, G., 2014. A proposal for an integrated modelling framework to characterise habitat pattern. Environmental Modelling \& Software 52, 176-191.
\goDOI{10.1016/j.envsoft.2013.10.011}

\medskip
\checkendpaper{}

%-------------------------------------
%%%%%%%% REFERENCES
%-------------------------------------

\begin{footnotesize}
\renewcommand*\labelenumi{[\theenumi]}
\raggedright
\nohyphens{

}
\end{footnotesize}

\begin{thebibliography}{}

\bibitem{Bennett_2010} Bennett, J., 2010. OpenStreetMap. Packt Publishing. ISBN: 978-1-84719-750-4

\bibitem{Bennett_etal_2010} Bennett, G., Bento Pais, R., Berry, P., Didicescu, P. S., Fichter, M., Hlav\'a\v{c}, V., Hoellen, K., Jones-Walters, L., Miko, L., Onida, M., Plesn\'ik, J., Smith, D., Wakenhut, F., 2010. Green Infrastructure Implementation: Proceedings of the European Commission Conference 19 November 2010. (Ed: Karhu, J.). European Commission, 28 pp.
\gourl{http://ec.europa.eu/environment/nature/ecosystems/green_infrastructure.htm}

\bibitem{de_Rigo_2012} de Rigo, D., 2012. Applying semantic constraints to array programming: the module ''check\_is'' of the Mastrave modelling library. In: Semantic Array Programming with Mastrave - Introduction to Semantic Computational Modelling. \gourl{http://mastrave.org/doc/mtv_m/check_is}

\bibitem{DGENV_2012a} Directorate-General~for~Environment~(DG ENV), 2012. The Multifunctionality of Green Infrastructure. Science for Environment Policy. \gourl{http://ec.europa.eu/environment/nature/ecosystems/docs/Green_Infrastructure.pdf}

\bibitem{DGENV_2012b} Directorate-General~for~Environment~(DG ENV), 2012. Natura 2000 data - the European network of protected sites. Temporal coverage: 2011. European Environment Agency web portal. \gourl{http://www.eea.europa.eu/data-and-maps/data/ds_resolveuid/60860bd4-28d6-44aa-93c7-d9354a8205e3}

\bibitem{EC_1992} European~Commission, 1992. Council directive 92/43/EEC of 21 may 1992 on the conservation of natural habitats and of wild fauna and flora. Official Journal of the European Union 35 (L 206), 7-50.

\bibitem{EC_2010} European~Commission, 2010. Directive 2009/147/EC of the European Parliament and of the Council of 30 November 2009 on the conservation of wild birds. Official Journal of the European Union 53 (L 20), 7-25.

\bibitem{EC_2011} European~Commission, 2011. Our life insurance, our natural capital: an EU biodiversity strategy to 2020. Brussel, COM (2011) 244 final. \gourl{http://ec.europa.eu/environment/nature/biodiversity/comm2006/pdf/2020/1_EN_ACT_part1_v7\%5B1\%5D.pdf}

\bibitem{EC_2013} European~Commission, 2013. Green Infrastructure (GI) - Enhancing Europe's Natural Capital. Brussel, COM (2013) 249 final. \gourl{http://eur-lex.europa.eu/LexUriServ/LexUriServ.do?uri=COM:2013:0249:FIN:EN:PDF}

\bibitem{EEA_2012} European~Environment~Agency, 2012a. Corine Land Cover 2006 raster data - version 16. European Environment Agency web portal. \gourl{http://www.eea.europa.eu/data-and-maps/data/ds_resolveuid/ef13cef8-2ef5-49ae-9545-9042457ce4c6}

\bibitem{Estreguil_etal_2012} Estreguil, C., Caudullo, G., de Rigo, D., Whitmore, C., San-Miguel-Ayanz, J., 2012. Reporting on European forest fragmentation: standardized indices and web map services. IEEE Earthzine 5 (2), 384031+. \gourl{http://www.earthzine.org/?p=384031} (2nd quarter theme: Forest Resource Information).

\bibitem{Estreguil_etal_2013} Estreguil, C., Caudullo, G., San-Miguel-Ayanz, J., 2013. Connectivity of Natura 2000 Forest Sites. EUR 26087EN. Luxemburg: Publications Office of the European Union. JRC 83104. \goDOI{10.2788/95065}

\bibitem{Estreguil_etal_2014} Estreguil, C., de Rigo, D., Caudullo, G., 2014. A proposal for an integrated modelling framework to characterise habitat pattern. Environmental Modelling \& Software 52, 176-191. \goDOI{10.1016/j.envsoft.2013.10.011}

\bibitem{Estreguil_etal_2014b} Estreguil, C., de Rigo, D., Caudullo, G., 2014. Supplementary materials for: A proposal for an integrated modelling framework to characterise habitat pattern. \gourl{http://mastrave.org/bib/Estreguil\_etal\_EMSsuppl\_2014.pdf} \\[1mm](Extended version of the supplementary materials as published in Environmental Modelling \& Software 52, 176-191, \goDOI{10.1016/j.envsoft.2013.10.011}).

\bibitem{Haklay_Weber_2008} Haklay, M., Weber, P., 2008. OpenStreetMap: User-Generated Street Maps. Pervasive Computing 7(4). doi: 10.1109/MPRV.2008.80. 

\bibitem{McHugh_Thompson_2011} McHugh, N., Thompson, S., 2011. A rapid ecological network assessment tool and its use in locating habitat extension areas in a changing landscape. Journal for Nature Conservation 19 (2011) 236-244. \goDOI{10.1016/j.jnc.2011.02.002}.

\bibitem{Saura_Torne_2009} Saura, S., Torn\'e, J., 2009. Conefor Sensinode 2.2: a software package for quantifying the importance of habitat patches for landscape connectivity. Environmental Modelling \& Software 24 (1), 135-139. \goDOI{10.1016/j.envsoft.2008.05.005}.

\bibitem{Soille_Vogt_2009} Soille, P., Vogt, P., 2009. Morphological segmentation of binary patterns. Pattern Recogn. Lett. 30 (4), 456-459. \goDOI{10.1016/j.patrec.2008.10.015}.

\bibitem{Van_Rossum_Drake_2011} Van Rossum, G., Drake Jr., F., 2011. The Python Language Reference Manual (version 3.2). Network Theory Limited, ISBN 978-1-906966-14-0.


\end{thebibliography}
\end{document}